\documentclass[a4,12pt]{article}

\begin{document}

\titlepage
\begin{flushright}{CERN-TH/2001-258\\
T01/103\\
hep-th/0109207 \\
}
\end{flushright}
\begin{center}
{\large \bf Quintessence Model Building}
\end{center}
\begin{center}
Ph. Brax\footnote{On leave of absence from Service de Physique Th\'eorique,
CEA-Saclay, F-91191, Gif/Yvette cedex, France}
\end{center}
\begin{center}
CERN, Theory Division, CH-1211 Geneva 23 , Switzerland.
\end{center}

\begin{center}
J. Martin
\end{center}
\begin{center}
Institut d'Astrophysique de Paris, 98 Boulevard Arago,
75014, Paris, France.
\end{center} 
\begin{center}
A. Riazuelo
\end{center}
\begin{center}
D\'epartement de Physique Th\'eorique, Universit\'e de
Gen\` eve, 24 Quai Anselmet 1211 Gen\`eve-4, Switzerland.
\end{center}

\vskip .1 cm
\begin{center}
{\bf Abstract}
\end{center}
\vskip .1 cm
\noindent A short review of some of the aspects of quintessence model building
is presented. We emphasize the role of tracking models and
their possible supersymmetric origin.

\section{Introduction}

The cosmological constant problem is one of the thorniest issues of
present day theoretical physics\cite{Wein}. It has recently received a
new twist from the analysis of the luminosity distance vs redshift
diagram of type Ia supernovae\cite{SNIa} combined with the cosmic
background anisotropy results\cite{B98a,B98b,MAX1a,MAX1b}. It appears
that our universe seems to be in a late accelerated phase of expansion
driven by a vacuum energy density of the order of the critical energy
density of the universe.  This is strongly suggested by the
observation of a negative acceleration parameter $q_0$ and the
location of the first Doppler peak, a result compatible with a
spatially flat universe in agreement with the inflationary scenario.

From a theoretical point of view this poses two highly different
cosmological constant problems.  The first one corresponds to the lack
of explanation for an almost exact vanishing of the cosmological
constant.  Indeed modern theories of particle physics predict that
radiative corrections and phase transitions such as the one occurring
at the weak scale, the scale of the electroweak symmetry breaking, or
even worse the Planck scale, contribute to the fourth power to the
vacuum energy. Without the need to invoke any putative physics beyond
the standard model of particle physics one can see that a term like
$M_W^4 \simeq 10^8 \hbox {GeV}^{4}$ destroys the delicate balance
required by the almost vanishing of the cosmological constant.  Within
the realm of high energy physics such an acute cancellation between
diverse sources of vacuum energy appears as a fine-tuning of the
parameters of the theory and requires a dynamical explanation. One of
the most commonly advocated mechanisms is the presence of a symmetry
which would naturally lead to the vanishing of the cosmological
constant\cite{Witten}.

Supersymmetry is such a symmetry\cite{fer}. As a matter of fact
globally supersymmetric theories have a vanishing vacuum
energy\cite{BS}. Unfortunately our world does not seem to be described
by a globally supersymmetric theory; despite intensive searches,
supersymmetric partners of the leptons such as the selectrons have
never been observed. This is usually accommodated by requiring that
supersymmetry is softly broken therefore losing the exact vanishing of
the cosmological constant. Moreover while incorporating gravity in the
supersymmetric setting, i.e. supergravity, one finds that
supersymmetric vacua do not automatically lead to a vanishing vacuum
energy. Supergravity vacua tend to favor a negative cosmological
constant and various no-go theorems forbid the existence of de Sitter
or accelerating supergravity vacua\cite{mal}. Similarly string
theory\cite{str} in its usual form cannot be formulated in de Sitter
or accelerating universes due to the presence of a horizon, and the
impossibility of defining a scattering S-matrix\cite{kal}.

The quintessence proposal does not address the previous hard
cosmological constant problem\cite{WCOS,Wett}. It tackles the issue
of obtaining a small and non-vanishing cosmological constant provided
that an unknown mechanism guarantees that the hard cosmological
constant problem is solved.  In practice this amounts to studying
field-theoretical models involving one or several fields evolving in
time under the combined action of gravity and field
interactions. Quintessence arises when the present value of the
potential is compatible with a fraction $\Omega_Q \simeq 0.7$ of the
critical energy density $\rho_{\rm c} \simeq 10^{-48} \hbox {GeV}^{4}$
of the universe. Notice that this involves finding models where the
vacuum energy of the universe is some hundred and twenty orders of
magnitude smaller than the Planck scale. Of course most models with
such a large hierarchy of scales will suffer from some kind of
fine-tuning.  Fortunately there exists a class of models, the tracking
models, where the quintessence field rolls down a potential while
converging towards an attractor\cite{RP}. This tracking mechanism
leads to small values of the vacuum energy without invoking a
fine-tuning of the initial conditions. Still it requires the choice of
a particular scale $M$ for the quintessence sector. Such a scale can
be chosen to be large, for instance $10^{10} \hbox {GeV}{}$, therefore
avoiding the need for an arbitrary small scale in the model. Of course
it would be nice if the choice of this quintessence scale could be
linked to some more fundamental scales such as the Planck scale or the
string scale.

As already mentioned the best possible candidates for solving the hard
cosmological constant problem is supergravity, or string
theory. Nevertheless it seems that finding accelerating supersymmetric
vacua in these theories is an impossible task. Of course this is not
the case anymore when supersymmetry is broken. One suggestive
possibility is that supersymmetry is broken while preserving a
vanishing cosmological constant, this could happen for instance as the
result of an accidental cancellation. In that setting one can try to
implement the quintessence proposal and look for quintessence models
with broken supersymmetry.  We will present some examples
here\cite{bm_99}.

In a first section we will recall the four distinct problems that
quintessence models should address. Then we will present examples of
simple quintessence models, such as exponential models. Eventually we
will introduce tracking models and discuss the issue of supersymmetric
tracking models\cite{bm_99,br,CNR}.

\section {Quintessential Problems}

In the following we will restrict ourselves to quintessence models
with a single quintessence field $Q$. More precisely the dynamics of
this scalar field will be governed by the Friedmann and Klein-Gordon
equations.  The pressure and energy density of a scalar field are
defined by
\begin{equation}
p_{Q} = \frac{\dot Q^2}{2}- V(Q),\ \rho_{Q} = \frac{\dot Q^2}{2}+ V(Q), 
\end{equation}
where $V(Q)$ is the potential energy of the quintessence field.  The
evolution of the scale factor of the universe $a$ is dictated by the
Friedmann equation
\begin{equation}
H^2\equiv \biggl(\frac{\dot{a}}{a}\biggr)^2 = \frac{\kappa}{3}\rho ,
\end{equation}
where $\rho$ is the total energy density involving cold dark matter,
radiation and the quintessence field.  Notice that we have assumed an
exactly vanishing cosmological constant.  The Klein-Gordon equation
gives the time evolution of the quintessence field
\begin{equation}
\ddot Q+3H \dot Q = -\frac{\partial V(Q)}{\partial Q}.
\end{equation}
The phenomenological constraint on $\rho_Q$ is that it should
correspond to an energy scale of order $5.7 \times 10^{-47}
\hbox{GeV}^{4}$ now.

\subsection{ The Coincidence problem}

The cosmological coincidence problem is the ``why now'' question, i.e.
why should we observe a slight dark energy dominance compared to the
other forms of matter now?  In the context of quintessence this can be
reformulated as follows. Why does the evolution of the quintessence
field leads to dominance now? In particular one should be wary of a
very serious initial conditions problem. Indeed it would be extremely
unnatural to assume that coincidence occurs due to a particular choice
of the initial conditions for $Q$. For that reason tracking
quintessence models are favored. They lead to the independence of
coincidence with respect to the initial conditions. This is due to the
presence of an attractor.

Another coincidence is that dark energy dominates soon after
matter-radiation equality.  This is often dubbed as the triple
coincidence problem.  Assuming that cold dark matter is due to a
stable particle with interactions governed by a scale $M_{\rm CDM}$ it
has been observed that the energy density at matter-radiation equality
scales as $M_{\rm CDM}^8/m_{\rm Pl}^4$\cite{hall}.  In order to have a
natural explanation for the triple cosmic coincidence this requires
that
\begin{equation}
\rho_Q \vert_{\rm now} \simeq \frac{M_{\rm CDM}^8}{m_{\rm Pl}^4},
\end{equation}
where $M_{\rm CDM} \simeq 10^3 \hbox{GeV}{}$.  This suggests that a
see-saw mechanism is at play.

\subsection{The Fine-Tuning Problem}

Any choice of potential does not lead to a valid quintessence
model. Indeed one should avoid unnatural choices of parameters. For
instance take the potential $ V(Q) = m^2Q^2/2$ where the only free
parameter must be chosen to be $m \simeq 10^{-33} \hbox {eV}{}$, a
scale as difficult to explain as a small vacuum energy.  As already
mentioned, fom tracking models, the potential depends on a single
scale $M$
\begin{equation}
V(Q) = \frac{M^{4+\alpha}}{Q^{\alpha}}.
\end{equation}
Depending on $\alpha$, the value of $M$ has to chosen in order to
obtain $\Omega_Q = 0.7$ now.  For instance for $\alpha = 6$ one finds $M = 
10^6 \hbox{GeV}{}$.  Observe that such a choice of $M$ can be viewed
as a fine-tuning, why should physics involve such an energy scale? On
the other hand one can argue that this does not involve any
fine-tuning as such an energy scale is compatible with high energy
physics.  Granted such a value for $M$ we are guaranteed that
coincidence occurs now in a way totally independent of the initial
conditions.

\subsection{The equation of state}

Another phenomenological constraint on quintessence models concerns
the equation of state
\begin{equation}
\omega_Q\equiv \frac{p_Q}{\rho_Q}.
\end{equation}
For a pure cosmological constant $\omega_{\Lambda} = -1$ while recent
data suggest $\omega_Q\le -0.8$. This is a very stringer restriction
on various types of models. Most models will fail to satisfy this
criterion.

\subsection{Model Building}

One would like to find natural quintessence models which can be
embedded in high energy physics. Eventually one would like to
understand quintessence as emerging from string theory.  One
interesting possibility is to consider the fate of moduli in string
theory, i.e. fields like the dilaton.  As a rule these fields have no
potential perturbatively while a non-vanishing runaway potential is
generated non-perturbatively.  It is tempting to think of the
quintessence field as one of these moduli. In the case of the dilaton
the potential is of the exponential of exponential type and does not
lead to an appropriate potential\cite{Bine}.  More work along these
lines will certainly be fruitful.

Another possibility is to try to link the quintessence field to the
breaking of supersymmetry. Of course this requires to parameterize the
spontaneous breaking of supersymmetry\cite{bri}. We will present such
an example.

\section{Some Quintessence Models}

In order to illustrate the previous considerations we will consider
two typical examples.  In particular we will pay attention to the
initial conditions problem.  This problem is solved
provided\cite{SWZ,ZWS}
\begin{equation}
\Gamma = \frac{V(Q)V''(Q)}{V'(Q)^2}>1
\end{equation}
during the evolution of $Q$. This guarantees the existence of an
attractor and the loss of sensitivity to the initial conditions.

\subsection{Pseudo Goldstone Boson Model}

Consider the potential which can arise for very light axions after the
breaking of the Peccei-Quinn symmetry\cite{fri}
\begin{equation}
V(Q) = M^4\biggl[\cos \biggl(\frac{Q}{f}\biggr ) +1\biggr].
\end{equation}
There are two scales in this potential. The scale $f$ is the symmetry
breaking scale which is chosen to be $f\ge 10^{16}
\hbox{GeV}{}$. Starting with $Q\ll f$ we see that $Q$ will roll down
towards the minimum of the potential. The scale $M$ must be chosen to
be small enough in order to reach $\Omega_Q = 0.7$ now. For instance
one can take $M \simeq 10^{-1} \hbox{eV}{}$. Of course this is a
fine-tuning which receives no explanation here and jeopardizes the
predictability of the model\footnote{See\cite{hall} for a model where
$M$ is related to the supersymmetry breaking scale in a hidden
sector.}. Moreover the initial conditions must be tuned to reach the
right vacuum energy now\cite{well}.  This can be seen by noticing that
$\Gamma$ becomes much smaller than one when $Q$ is close to the
minimum of the potential.

\subsection {Exponential Potentials}

These potentials have the nice feature of being naturally generated by
compactifying higher dimensional supergravities\cite{tow}.  The
potential
\begin{equation}
V(Q) = V_0e^{-\lambda Q}
\end{equation}
depends on one scale $V_0$ and a constant $\lambda$.  Due to the
exponential behavior large scale hierarchies can be easily generated
for a high value of $V_0$. For a large range of initial conditions
this model possesses an attractor which tracks the behavior of the
dominant matter or radiation energy density\cite{FJ1}.  Unfortunately
such an attractor leads to an effective equation of state for the
quintessence field which coincides with the one of the tracked matter
density. At late times this means $\omega_Q = 0$ and therefore no
acceleration of the universe. This can be remedied by choosing the
initial conditions outside of the basin of attraction of the
attractor. This has the unwanted feature of leading to a dependence on
the initial conditions.  The initial conditions problem can be
remedied by double exponential potentials.

\section{Tracking Quintessence Models}

These models preserve the nice tracking property of exponential
potentials while leading to a negative equation of state and late
acceleration.  As $\Gamma>1$ the late evolution of the quintessence
field is independent of the initial conditions. The quintessence field
reaches an attractor when either radiation or matter dominates and
leaves the attractor in the recent past to start dominating over the
CDM matter density. The only parameter $M$ has to be tuned in order to
reach the appropriate value of $\Omega_Q$ now.

These models can be obtained in a supersymmetric context by
considering a super-Yang-Mills theory with gauge group $SU(N_c)$ and
$N_F<N_c$ quark flavors\cite{Bine}. At energies below the gluino
condensation scale $M$ a non-perturbative superpotential for the meson
fields is generated. It leads to a inverse power law potential.  In
this model a large condensation scale like $10^{6} \hbox {GeV}{}$ for
$\alpha = 6$ is not unnatural.

These models suffer from two main problems. On the one hand the first
problem springs from the observation that they lead to an equation of
state $\omega_Q$ which is always larger that $-0.7$, i.e. falling
outside of the experimental ball-park.

On the other hand the supersymmetric tracking model is not consistent
unless one takes supergravity into account\cite{bm_99}. Indeed this
can be easily seen by noticing that on the attractor, and therefore
just before coincidence, one finds
\begin{equation}
V'' = \frac{9}{2}\frac{\alpha +1}{\alpha}(1-\omega_Q^2)H^2,
\label{mass}
\end{equation}
where $V''$ is the second derivative of the potential.
Using Friedmann's equation, this leads to
\begin{equation}
Q\vert_{\rm now} \simeq m_{\rm Pl}.
\end{equation}
A globally supersymmetric theory receives supergravity corrections in
the form of $Q/m_{\rm Pl}$ which cannot be neglected in the late stage
of the evolution of the universe. In particular a naive calculation of
the supergravity potential for the gaugino condensation
model\footnote{We assume a flat K\"ahler potential.}  leads to a
negative energy density in the recent past.

We therefore conclude that the description of quintessence by tracking
fields requires an embedding of the model in supergravity where terms
in $Q/m_{\rm Pl}$ are properly taken into account.

\section{Supergravity Tracking Models}
\subsection{ The Supergravity Tracking Potential}

We consider the possibility that tracking models arise from
supergravity in four dimensions. Eventually these models must be
considered as an effective field theory description of a more
fundamental theory. We will therefore only concentrate on describing
mechanisms leading to tracking potentials in four dimensional
supergravity leaving an explanation for the origin of these models to
more refined studies.  As a rule the potential in supergravity depends
on a single function $G = \kappa K +\ln (\kappa^3 \vert W\vert^2)$
where $K$ is the K\"ahler potential governing the kinetic terms
$K_{i\bar j}\partial_{\mu}\phi^i\partial^{\mu}\bar \phi^{\bar j}$and
$W$ is the superpotential. The scalar potential reads
\begin{equation}
V(Q) = \kappa^{-1}e^G(G^iG_i -3) +V_{\rm D},
\end{equation}
where $V_{\rm D}$ is positive definite and involves the gauge sector
of theory. In the following we consider flat $D$ directions where
$V_{\rm D}$ vanishes altogether. Globally supersymmetric theories can
be obtained by letting $\kappa\to 0$. In that case the potential
becomes positive definite. Consider the case of quintessence with a
low value of the initial condition for $Q$. In that case supergravity
corrections are small and one can trust the globally supersymmetric
approximation. Assume that the models leads to an inverse power law
model for low values of $Q$. As $Q$ evolves towards the Planck scale,
the main supergravity correction will come from the prefactor
$e^{\kappa K}$ which leads to the supergravity corrected
potential\cite{bm_99}
\begin{equation}
V(Q) = e^{\frac{\kappa}{2}Q^2}\frac{M^{4+\alpha}}{Q^{\alpha}}.
\end{equation}
Notice that the exponential correction does not play a role for most
of the evolution of $Q$. It only becomes relevant when the
quintessence field dominates leading to a rapid increase of the
potential energy. This has the nice phenomenological feature of
leading to an equation of state $\omega_Q \simeq -0.82$ almost
independently of $\alpha$.
\subsection{A Supergravity Model}

The construction of supergravity quintessence models necessitates the
existence of three independent sectors\footnote{We acknowledge an
interesting discussion with P. Bin\'etruy on that issue.}, the
observable sector where the standard model fields live, a supergravity
breaking sector where the superpartners acquire their masses and a
quintessence sector where supersymmetry is also broken\cite{br}. The
three independent sectors can be coupled by gravity for instance. The
quintessence sector must be decoupled from the observable sector as
the effective mass of the quintessence field now, see
Eq.~(\ref{mass}), is of the order of the Hubble parameter $H_0 \simeq
10^{-43} \hbox{GeV}{}$. Such a small mass would lead to the existence
of a long range fifth force contradicting experimental
facts\cite{mas}. The quintessence sector and the supersymmetry
breaking sectors must be decoupled to prevent the existence of time
dependent sparticle masses\cite{br}. If time-dependent sparticle
masses are allowed the large contribution of the quintessence field
$Q$ to the supersymmetry breaking $F$-terms might lead to very large
values of the sparticle masses, preventing any possible detection. We
discard this possibility here.

Let us now consider a typical string inspired scenario involving type
I models\cite{ak}. Such models comprise branes where interactions are
set by the string scale $m_{\rm S}$ and suppressed gravity
effects\cite{rigo}.  Moreover they contain many anomalous $U(1)$ gauge
groups with associated Fayet-Iliopoulos terms\cite{dw} of the order
of the string scale\footnote{ These are moduli associated to the
blowing up modes of the orbifold compactification. We assume that they
are somehow stabilized at a value of the order of the string scale.}.

At low energy the corresponding (broken) supergravity models will
possess two characteristic scales, the string scale $m_{\rm S}$ that
we suppose to be much smaller than the Planck scale.  We can construct
a simple supergravity quintessence model by using three fields, the
quintessence field $Q$ and two fields $X$ and $Y$ of charges $1$ and
$-2$ under an anomalous $U(1)$ gauge group. We assume that the leading
interaction between $X$ and $Y$ is given by the superpotential
$W = X^2Y$. The model possesses a D-flat direction where $Y = 0$ and
$X = m_{\rm S}$ due to the presence of a non-vanishing Fayet-Iliopoulos
term.  Moreover we assume that the K\"ahler potential takes the form
\begin{equation}
K = Q\bar Q + Y\bar Y f(Q,\bar Q, m_{\rm S}, m_{\rm Pl}) + X\bar X.
\end{equation}
Notice that we have a non-trivial coupling between $Y$ and $Q$ in the
K\"ahler potential. Due to the large hierarchy between the string and
Planck scales we take $f(Q, \bar Q, m_{\rm S},m_{\rm Pl})\equiv
f(Q\bar Q/m_{\rm S}^2)$ to be a positive increasing function
normalized by $f(0) = 1$. For $Q\ll m_{\rm S}$ we can expand $f$ in
power series. For larger values non-perturbative information might be
necessary. We assume that $f$ can always be defined by a polynomial of
degree $p$. It is now an easy exercise to compute the scalar potential
\begin{equation}
V(Q) = e^{\kappa Q^2/2}m_{\rm S}^4f^{-1}\biggl(\frac{Q\bar Q}{m_{\rm S}^2}\biggr ).
\label{pot}
\end{equation}
For small values of $Q\ll m_{\rm S}$ the potential is almost constant
$V(Q) \simeq m_{\rm S}^4$. For larger values of $Q$ the potential
decreases before reaching large values of $Q$ where the exponential
supergravity correction becomes relevant.  Notice that the potential
energy is always smaller than $m_{\rm S}^4$ as necessary for the low
energy effective action approach to be valid.

Assuming that $f$ is polynomial at large $Q$ leads to
\begin{equation}
V(Q) \simeq e^{\frac{\kappa}{2}Q^2} \frac{m_{\rm S}^{4+2p}}{Q^{2p}},
\end{equation}
where we have chosen $Q = \bar Q$\footnote{See the related talk by
R. Caldwell for a discussion of spinessence.}.  Notice that this is
the supergravity tracking potential with $\alpha = 2p$.  So by allowing
$Q$ to roll down the potential (\ref{pot}) we find that the
quintessence potential behaves effectively like a tracking potential
before feeling the effect of the supergravity corrections.  Moreover
choosing the string scale to be of order of the TeV scale leads to
$\alpha = 4$. Putting $Q \simeq m_{\rm Pl}$ now we find that
\begin{equation}
\rho_Q\vert_{\rm now} = \frac{m_{\rm S}^8}{m_{\rm Pl}^4},
\end{equation}
as suggested in order to solve the triple coincidence problem with
$M_{\rm CDM} = m_{\rm S} \simeq 10^3 \hbox {GeV}{}$.

In conclusion we have described a simple string inspired supergravity
model leading to a tracking model with interesting phenomenological
features. Of course much work remains to be done to make this picture
more explicit.

\section{ Conclusion}

We have presented some of the aspects of quintessence model building
putting the emphasis on the necessity of constructing models embedded
in high energy theories such as supergravity.  The subject of
quintessence model building is still in its infancy and is likely to
receive new impetus from string theory and its various avatars.

\section*{Acknowledgments}

One of us (P.B.) thanks the organizers of the conference for kindly
allowing him to present this work.

\end{document}